\documentclass[aps,preprint,showpacs]{revtex4}
\usepackage{graphicx}
\usepackage{amsmath}
\usepackage{epsfig}

\begin{document}

\title{Isotope effect and bond-stretching phonon anomaly in high-Tc cuprates}
\author{S. Cojocaru$^{1,2}$, R. Citro$^{3,4}$ and M. Marinaro$^{3,5}$}
\affiliation{$^{1}$ National Institute of Physics and Nuclear Engineering,
Bucharest-Magurele 077125, Romania}
\affiliation{$^{2}$ Institute of Applied Physics, Chi\c{s}in\u{a}u 2028, R. Moldova}
\affiliation{$^{3}$ Dipartimento di Fisica \textquotedblleft E. R.
Caianiello\textquotedblright\ and C.N.I.S.M., Universit{\`{a}} degli Studi
di Salerno, Baronissi (Sa)I-84081, Italy}
\affiliation{$^{4}$ Laboratorio SuperMat, CNR-INFM, Salerno, Italy}
\affiliation{$^{5}$ I.I.A.S.S., Via G. Pellegrino 19, Vietri sul Mare (SA) 84019, Italy}
\pacs{74.72.-h, 71.38.-k, 73.20.Mf}

\begin{abstract}
We analyse a model where the anomalies of the bond-stretching LO phonon mode
are caused by the coupling to electron dynamic response in the form of a
damped oscillator and explore the possibility to reconstruct the spectrum of
the latter from the phonon measurements. Preliminary estimates point to its
location in the mid infrared region and we show how the required additional
information can be extracted from the oxygen isotope effect on the phonon
spectrum. The model predicts a significant measurable deviation from the
"standard value" of the isotope effect even if the phonon frequency is far
below the electron excitation spectrum, provided the latter is strongly
incoherent. In this regime, which corresponds to the "mid infrared
scenario", the phonon linewidth becomes a sensitive and informative probe of
the isotope effect.
\end{abstract}

\maketitle

\section{Introduction}

One of the reasons the lattice degrees of freedom of cuprate superconductors
remain an intensively debated topic on both experimental and theoretical
sides is the existence of strong anomalies in some of the phonon spectrum
branches and their possible relation to electronic properties \cite{graf,
astuto, Reznik,Braden}. In particular, the high frequency, about $80$ meV,
in-plane bond-stretching LO ( BS or the "half-breathing") mode shows large
damping, typically several meV, and dispersion dip of about $20\%$ for
momenta $q_{x}$ ($q_{y}$) around $0.25$ r.l.u. The line broadening and
softening increase with hole doping and are sensitive to temperature close
to ${T}_{c}$  (see \cite{pint} for a review; the superconductivity induced 
anomalies have been discussed in \cite{Misochko} for Raman active
phonons). The first measurements of the Cu-O BS mode only reported a
cosine-like dispersion, which can be understood in terms of conventional
calculations. Subsequent measurements, however, pointed to anomalous
behavior related to charge inhomogeneities and different interpretations
have been given. McQueeney and co-workers \cite{McQ} reported a dynamic
doubling of the unit cell in the $Cu-O$ bond direction and discontinuous
phonon dispersion. This has been related to charge stripes with a slower
dynamics than the BS phonon frequency and formation of a second phonon mode
with a lower frequency. In contrast, Pintschovius and Braden \cite{PB} have
found a smooth BS mode dispersion with an anomalously large slope and
broadening around $q=(0.25,0,0)$. Reznik and co-workers \cite{Reznik1} have
studied $La_{2-x}Ba_{x}CuO_{4}$ for $x\sim 1/8,$ where quasistatic stripes
exist, and have assigned the observed strong enhancement of the phonon
broadening to the sharp softening. D'Astuto et al. \cite{dastuto} have
carried out high precision X-ray scattering experiments on $%
La_{2-x}Sr_{x}CuO_{4+\delta }$ and concluded that this phonon broadening is
caused by intrinsic damping and not to an apparent one due to steep
dispersion.

In an earlier paper \cite{ccm}, based on a semi-phenomenological model, we
have suggested that the anomalously strong softening and broadening of the
half-breathing mode could be related to the other anomaly observed in the
mid-infrared region of the optical conductivity spectra (MIR) in high-Tc
cuprates \cite{basov}. Apart from making this connection, the way our model
addresses the problem of the phonon anomaly is opposite to other approaches:
phonon properties are taken as an experimental input which then serves to
determine certain properties of the electron excitations. It explains that
the measurements taken so far are insufficient and motivates new
experiments. Namely, to bridge the existing gap it is important to know the
effect of oxygen isotope substitution upon the phonon spectrum. In what
follows we describe the details of this approach which can be directly
applied to the analysis of such experiments as well as to test the
consistency of the model. We also calculate the expected values of the
measurable parameters depending on the frequency of the electronic
oscillator which spans the range between MIR and near resonance frequencies.
As the BS phonon mode couples to electron density excitation only at $q\neq 0
$, it could give a unique opportunity to access the finite momentum
evolution of the MIR excitation. On the other side, the present work
could stimulate further theoretical investigation of the isotope effect
within microscopic models, as for instance on the mechanism of dynamical
stripe fluctuations \cite{Kaneshita}. An interesting possibility is
implied by the analysis of site-selective isotope substitution on Raman
active phonons \cite{Sherman} when, in contrast to total substitution, a
significant change of coupling occurs due to modification of phonon
eigenvectors.

\section{The model}

Description of the phonon softening relies on the Dyson equation for Green's
function (GF):%
\begin{equation}
D\left( q,\omega \right) =\frac{2\omega _{0}}{\omega ^{2}-\omega
_{0}^{2}\left( 1+\lambda \sin ^{2}\left( q_{x}\right) P\left( q,\omega
\right) \right) },  \label{Phonon}
\end{equation}%
where $\lambda $ is the dimensionless coupling constant and $P\left(
q,\omega \right) $ electron density response function \cite{Khaliullin}. It
is assumed that the part of this response which is relevant for the phonon
softening can be considered in the form of a damped oscillator \cite{ccm}
\begin{equation}
P\left( q,\omega \right) =\frac{\eta _{q}}{\omega ^{2}-\Omega
_{q}^{2}+i\Gamma _{q}\omega }.  \label{mode}
\end{equation}%
The oscillator is parametrised by the full-width at half-maximum (FWHM) $%
\Gamma _{q},$ frequency $\ \Omega _{q}$ and oscillator strength $\eta _{q}$
which have to be determined by analyzing the data on the anomalous BS phonon
branch. The experimentally determined values of the linewidth $\tilde{\gamma}%
_{q}$ and frequency $\tilde{\omega}_{q}$ are used to define the standard
form of the renormalized phonon GF%
\begin{equation*}
D^{-1}\left( q,\omega \right) \sim \omega ^{2}-\tilde{\omega}_{q}^{2}+i%
\tilde{\gamma}_{q}\tilde{\omega}_{q}.
\end{equation*}%
In our model the phonon GF is found by solving for the complex poles of (\ref%
{Phonon}) $\omega =\omega _{q}-i\gamma _{q}:$%
\begin{equation*}
\omega _{0}^{2}-\omega _{q}^{2}+\gamma _{q}^{2}=\frac{\xi _{q}\omega
_{0}^{2}\ \left( \Omega _{q}^{2}-\Gamma _{q}\gamma _{q}-\omega
_{q}^{2}+\gamma _{q}^{2}\right) }{\left( \Omega _{q}^{2}-\Gamma _{q}\gamma
_{q}-\omega _{q}^{2}+\gamma _{q}^{2}\right) ^{2}+\left( \left( \Gamma
_{q}-2\gamma _{q}\right) \omega _{q}\right) ^{2}},
\end{equation*}%
\begin{equation}
\ 2\gamma _{q}=\frac{\xi _{q}\omega _{0}^{2}\ \left( \Gamma _{q}-2\gamma
_{q}\right) }{\left( \Omega _{q}^{2}-\Gamma _{q}\gamma _{q}-\omega
_{q}^{2}+\gamma _{q}^{2}\right) ^{2}+\left( \left( \Gamma _{q}-2\gamma
_{q}\right) \omega _{q}\right) ^{2}}\allowbreak .  \label{dispersion}
\end{equation}%
Thus, we have the correspondence $\tilde{\omega}_{q}=\sqrt{\omega
_{q}^{2}-4\gamma _{q}^{2}}$ and $\tilde{\gamma}_{q}\tilde{\omega}%
_{q}=2\omega _{q}\gamma _{q},$where the phonon linewidth measured in
experiments, $\tilde{\gamma}_{q},$ is roughly twice as large as the quantity
we use in the calculations, $\gamma _{q}.$ Still these two equations are
insufficient to determine the three parameters of the model, $\Gamma
_{q},\Omega _{q}$ and $\xi _{q}=\lambda \eta _{q}\sin ^{2}\left(
q_{x}\right) $, and we further discuss additional input that could be
obtained by varying the frequency of the phonon mode, i.e., by oxygen
isotope substitution $O^{16}\rightarrow O^{18}$.

In fact, there exists a detailed information about the isotope effect (IE)
on electronic properties of cuprates, e.g. on $T_{c}$ \cite{pringle} and
ARPES \cite{iwasawa}, but there have apparently been no attempts to consider
the effect such substitution has on the phonon spectra. It is known that the
high energy phonon modes, and in particular the one considered here, are
dominated by oscillations of oxygen ions. Therefore such experiments should
contain important information about both lattice and electronic excitations.
The mass dependence of the phonon dispersion has been studied for other
materials. For instance in \cite{simonelli}, where suppressed boron IE on T$%
_{c}$ of the $MgB_{2}$ is related to giant anharmonicity and nonlinear
electron-phonon coupling \cite{yildirim}. In cuprates anharmonicity could be
relevant for apical oxygen \cite{crespi}, but for the in plane modes the
harmonic approximation was shown to be accurate \cite{falter} and is
therefore adopted for the present model. Moreover, as discussed in \cite{gun}%
, the width of the half-breathing phonon in cuprates is mainly due to the
electron--phonon coupling. In more conventional materials, like $\alpha -Sn$
\cite{wang} or $Ge$ \cite{gobel}, the isotopic frequency shifts show a
dependence inversely proportional to the square root of the mass and a
linewidth inversely proportional to the mass. The error of such measurements
is within a fraction of $cm^{-1},$ the isotopic effect on the linewidth is
of the order of a few $cm^{-1}$ and is due to anharmonicity and isotopic
disorder. Since in cuprates the linewidth of the anomalous phonon is an
order of magnitude larger, one might expect that also the IE should be
enhanced respectively. However, it will be shown below that the answer is
not so straightforward.

The quantity $\xi _{q}$ in (\ref{dispersion}) contains the coupling
constant, the structure factor of the BS phonon and the spectral weight of
the response function (\ref{mode}). It can be trivially eliminated to obtain
the first equation of the required set:%
\begin{equation}
\frac{2\gamma _{q}}{\Gamma _{q}}=\frac{\omega _{0}^{2}-\omega
_{q}^{2}+3\gamma _{q}^{2}}{\Omega _{q}^{2}-\left( 2\omega _{q}^{2}-2\gamma
_{q}^{2}-\omega _{0}^{2}\right) }.  \label{A}
\end{equation}%
Eq.(\ref{A}) gives a simple approximate relation between softening and
linewidth, generally a relatively small quantity: $\gamma _{q}\sim
C_{q}\left( \omega _{0}-\omega _{q}\right) .$ It is useful when analyzing
phonon spectra and making analytic estimates below. The other two equations
can be obtained from (\ref{dispersion}) by calculating the isotope
coefficients (IC) according to their usual definitions%
\begin{equation}
\alpha _{q}=-\frac{d\ln \omega _{q}}{d\ln M},\ \ \ \ \ \ \beta _{q}=-\frac{%
d\ln \gamma _{q}}{d\ln M}.\   \label{BC}
\end{equation}%
We then get simple but somehow lengthy expressions for the two momentum
dependent quantities (below the momentum index is omitted for brevity) that
have to be obtained from the proposed experiments:%
\begin{equation}
a=\alpha -0.5,\ \ \ \ b=\beta -1;  \label{def}
\end{equation}%
\begin{equation*}
a=\frac{v_{2}u_{12}+v_{1}u_{22}}{u_{21}u_{12}+u_{22}u_{11}},\ \ \ \ \ b=%
\frac{v_{2}u_{11}-v_{1}u_{21}}{u_{21}u_{12}+u_{22}u_{11}};
\end{equation*}%
where%
\begin{eqnarray*}
u_{11} &=&\frac{\left( 1-g^{2}/2+gy-x^{2}-y^{2}\right) \left( 4xy\right) ^{2}%
}{\left( g-2y\right) ^{2}\left( \psi ^{2}-x^{2}+y^{2}\right) ^{2}+\left(
2xy\right) ^{2}}, \\
u_{12} &=&\left( 1+\frac{2y}{\left( g-2y\right) }-\frac{\left( 2y\right)
^{3}\left( 1-gy+x^{2}+y^{2}\right) }{\left( g-2y\right) \left( \left( \psi
^{2}-x^{2}+y^{2}\right) ^{2}+\left( 2yx\right) ^{2}\right) }\right) , \\
v_{1} &=&\frac{2y}{\left( g-2y\right) }-\frac{8\left( yx\right) ^{2}\left(
1-g^{2}/2+gy-x^{2}-y^{2}\right) }{\left( g-2y\right) ^{2}\left( \left( \psi
^{2}-x^{2}+y^{2}\right) ^{2}+\left( 2yx\right) ^{2}\right) }-\frac{\left(
2y\right) ^{3}\left( 1-gy+x^{2}+y^{2}\right) }{\left( g-2y\right) \left(
\left( \psi ^{2}-x^{2}+y^{2}\right) ^{2}+\left( 2yx\right) ^{2}\right) }, \\
u_{21} &=&-\left( 2x^{2}\right) \left( \frac{1-2y/\left( g-2y\right) }{\psi
^{2}-x^{2}+y^{2}}+\frac{\left( 2y\right) ^{2}}{\left( g-2y\right) ^{2}}\frac{%
2\left( 1+gy-x^{2}-y^{2}\right) -g^{2}}{\left( \left( \psi
^{2}-x^{2}+y^{2}\right) ^{2}+\left( 2yx\right) ^{2}\right) }\right) , \\
u_{22} &=&\left( \frac{\left( 2y\right) ^{2}}{\psi ^{2}-x^{2}+y^{2}}-\frac{%
\left( 2y\right) ^{3}}{\left( g-2y\right) }\frac{\left(
1-gy+x^{2}+y^{2}\right) }{\left( \psi ^{2}-x^{2}+y^{2}\right) ^{2}+\left(
2yx\right) ^{2}}\right) , \\
v_{2} &=&\frac{\left( 2y\right) ^{3}}{\left( g-2y\right) }\frac{\left(
1-gy+x^{2}+y^{2}\right) }{\left( \psi ^{2}-x^{2}+y^{2}\right) ^{2}+\left(
2yx\right) ^{2}}-\frac{3y^{2}}{\psi ^{2}-x^{2}+y^{2}}+\frac{\left( 2y\right)
^{2}x^{2}}{\left( g-2y\right) ^{2}}\frac{2\left( 1+gy-x^{2}-y^{2}\right)
-g^{2}}{\left( \psi ^{2}-x^{2}+y^{2}\right) ^{2}+\left( 2yx\right) ^{2}} \\
&&-\frac{1}{\left( g-2y\right) }\frac{2yx^{2}}{\psi ^{2}-x^{2}+y^{2}}.
\end{eqnarray*}%
All the quantities above have been scaled with ~$\Omega _{q}$ ( Eq. (\ref{A}%
) can be scaled in the same way):
\begin{equation}
x_{q}=\frac{\omega _{q}}{\Omega _{q}},\ \ \ \ y_{q}=\frac{\gamma _{q}}{%
\Omega _{q}},\ \ \ \ \psi _{q}=\frac{\omega _{0}}{\Omega _{q}},\ \ \ g_{q}=%
\frac{\Gamma _{q}}{\Omega _{q}}.\   \label{scaled}
\end{equation}
The coefficients $a_{q}$ and $b_{q}$ define the deviations from the
reference values which are observed in many materials where such IE has been
measured. It is easy to see from (\ref{dispersion}) that for a large energy
separation between the phonon and electron excitations ($\Omega _{q}\gg
\omega _{q}$) one obtains $\alpha _{q}^{0}=0.5$ and $\beta _{q}^{0}=1$ given
that $\omega _{0}\sim M^{-1/2}:$ these are the "standard" values \cite{allen}
for the isotope coefficients (IC). To be mentioned that there is also a
purely phononic source of the "normal" value $\beta _{q},$ since the
linewidth of a phonon in isotopically pure perfect crystal is caused by
anharmonicity, which scales as $\gamma \sim M^{-1}$ at low temperature \cite%
{wang}.

When the energy of electronic excitation approaches the phonon frequency one
could expect these deviations should grow in absolute value, and we indeed
find this trend in the present model. However, there is a qualitative
difference between the linewidth IC $b_{q}$ and the dispersion IC, $a_{q}.$%
To clarify this point we show the explicit solution approximated by the
first few relevant terms when $\Omega _{q}$ and $\Gamma _{q}$ are the
largest parameters in the problem (actual calculations below were carried
out for the complete equations). Consequently, the linewidth $\gamma _{q}$
and the two coefficients are small and one can obtain an analytic estimate:%
\begin{equation*}
a\simeq y\left( 2g^{-1}x^{2}+g+2g\left( y/x\right)
^{2}-y/x^{2}-g^{-1}\right) ,
\end{equation*}%
\begin{equation}
b\simeq 2yg\left( 1-g^{-2}\right) +2x^{2}\left( 1-g^{2}/2\right) ,
\label{approx}
\end{equation}%
where we have again dropped the $q-$index. We see that $a_{q}$ vanishes with
phonon coupling constant $\xi _{q}$ (or with the linewidth $y_{q},$ see Eq. (%
\ref{dispersion})), while the coefficient $b_{q}$ remains finite and depends
on the energy separation between the two coupled excitations. Of course,
when $\xi _{q}\rightarrow 0$, one can not measure the mass shift of $\gamma
_{q}$, instead this relation indicates that for a vanishing linewidth one
should expect a saturation of the respective IC at a finite value determined
by the ratios like $x_{q}$ and $g_{q}$, while the dispersion IC is at its
"standard" value. Below we will see that there is another important property
of the $b_{q}$ coefficient, that it can take relatively large values even
when the two interacting excitations are far from resonance. Here we note
that the sign of the linewidth-derived IC depends on the ratio between $%
\Omega _{q}$ and $\Gamma _{q}.$

The set of three equations ((\ref{A}) and (\ref{BC}) or the approximation (%
\ref{A}) and (\ref{approx})) on the two parameters, $\Gamma _{q}$ and $%
\Omega _{q}$, is overcomplete. It therefore gives a possibility to check
also the consistency of the model when the respective data become available.
We now solve these equations for $a_{q}$ and $b_{q}$ with $x_{q}$ and $y_{q}$
as independent variables to demonstrate the correlation between the two
sets, but bearing in mind that $\omega _{q}$ and $\gamma _{q}$ are in fact
intrinsically connected through a microscopic mechanism of electron-phonon
interaction and it is the IC which are measured in experiments to localize
the "physical" point in $\left( x_{q},y_{q}\right) $. We use the data
obtained from inelastic neutron and X-ray scattering for some representative
values of the momentum transfer $q$ in $La_{1.85}Sr_{0.15}CuO_{4}$ (e.g.
\cite{astuto, Reznik}) and calculate these coefficients together with the
corresponding ratio $g_{q}$ for several assumed values of the electronic
frequency $\Omega _{q}.$ At the $\Gamma -$point the frequency of the BS
phonon is $\omega _{0}=85.5$ meV. For momentum 1) $q_{1}=\left(
0.25,0,0\right) $ we have $\tilde{\omega}_{q1}=73$ meV and $\tilde{\gamma}%
_{q1}=13$ meV; 2) for $q_{2}=\left( 0.22,0,0\right) $ we have $\tilde{\omega}%
_{q2}=78$ meV and $\tilde{\gamma}_{q2}=9$ meV; and 3) for $q_{3}=\left(
0.2,0,0\right) $ respectively $\tilde{\omega}_{q3}=81$ meV and $\tilde{\gamma%
}_{q3}=6.5$ meV. The results are shown in the table.

\begin{center}
\begin{tabular}{|c|c|c|c|c|}
\hline
$q$ (rlu) & assumed $\Omega $ (meV) & $b$ & $a$ & $g$ \\ \hline
$0.25$ & $400$ & $-0.074$ & $0.03$ & $2.4$ \\ \hline
$0.25$ & $300$ & $-0.025$ & $0.025$ & $1.78$ \\ \hline
$0.25$ & $150$ & $0.33$ & $-0.017$ & $0.79$ \\ \hline
$0.25$ & $100$ & $0.48$ & $-0.18$ & $0.36$ \\ \hline
$0.22$ & $400$ & $-0.15$ & $0.0278$ & $2.7$ \\ \hline
$0.22$ & $300$ & $-0.095$ & $0.021$ & $1.97$ \\ \hline
$0.22$ & $100$ & $1.7$ & $-0.09$ & $0.36$ \\ \hline
$0.22$ & $150$ & $0.37$ & $-0.002$ & $0.82$ \\ \hline
$0.2$ & $300$ & $-0.21$ & $0.018$ & $2.3$ \\ \hline
$0.2$ & $150$ & $0.29$ & $0.009$ & $0.9$ \\ \hline
\end{tabular}
\end{center}

The main trends are also illustrated by the Figs. 1 and 2.
\begin{figure}[tbph]
\centering
\includegraphics[height=5cm,width=6cm]{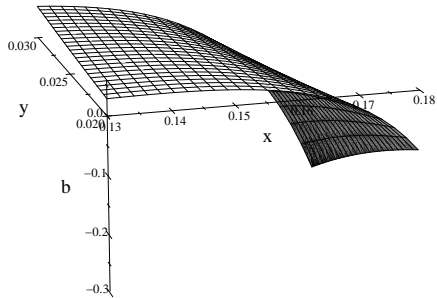}
\caption{Linewidth isotope coefficient $b_{q}$ calculated for $\Omega
_{q}=400$ meV in a broader range of $x$ $(=\protect\omega _{q}/\Omega _{q})$
and $y\left( =\protect\gamma _{q}/\Omega _{q}\right).$}
\label{fig1}
\end{figure}
As mentioned above, not all the values in the $\left( x,y\right) $ plane are
physically relevant and we have restricted the interval to the neighborhood
of the data used in the table. The dispersion IC is relatively small: it
varies within a few percents around the "standard" value $\alpha =0.5$ even
when $\Omega _{q} $ is close to the phonon frequency. Maximum of $a_{q}$ is
reached at $q=0.25, $ the momentum corresponding to the largest softening
and damping. In contrast, the IE on the phonon linewidth ($b_{q}$) can be
large even when the phonon and electron energy scales are rather distant ( $%
20\%$ and more for the "MIR scenario"), provided the electron excitation is
overdamped ($g>1 $). In this case $b_{q}$ is negative and $a_{q}$ is
positive. With "standard" $\beta _{q}=1$ this means a decrease below $\beta
_{q}=0.8$: as mentioned in relation to Eq.(\ref{approx}) this coefficient
tends to be negative when $\Gamma _{q}>\Omega _{q}$ and positive otherwise.
\begin{figure}[tbph]
\centering
\includegraphics[height=5cm,width=6cm]{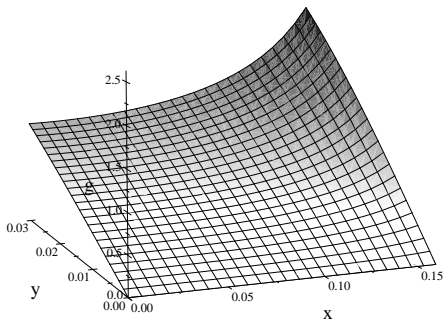}
\caption{Linewidth to frequency ratio $\Gamma _{q}/\Omega _{q}$ of the
electronic oscillator (\protect\ref{mode}) associated to the calculation in
Fig. 1.}
\label{fig2}
\end{figure}
Interestingly, that the sign of $a_{q}$ seems to have an opposite tendency.
For a given frequency $\Omega _{q}$ the largest values of the IC $b_{q}$ are
achieved in the region with maximal slope of $\omega $ ( $\gamma $ ) and not
where softening (linewidth) has reached its highest value. This is because
unlike $a_{q},$ which mainly depends on the value of linewidth (proportional
to softening), $b_{q}$ also strongly depends on the frequency ratio $\omega
_{q}/\Omega _{q}.$

From the above analysis it can be concluded that both isotope coefficients
of the half-breathing phonon mode contain a reach and complementary
information on the electron excitation responsible for the phonon anomalies.
This information can be used to extract the momentum resolved dynamical
electron density response function after a detailed experimental data on the
isotope coefficients become available. The response function is expected to
be strongly inhomogeneous and anisotropic following the strong momentum
dependence of the BS phonon softening. In our model it is fully described by
the three momentum dependent parameters of the damped oscillator (\ref{mode}%
). If the value of $b_{q}$ is found to be negative (or $\beta _{q}<1$), that
would be a strong indication in favor of the "MIR scenario" and would allow
to study the finite momentum counterpart of the anomaly known from optical
experiments by using the BS phonon as a probe. The accuracy of neutron and
X-ray scattering experiments is sufficient to quantify the oxygen isotope
effect, although it is "usually" weak. We have, however, shown that for
high-Tc cuprates the IE effect on the linewidth of the BS phonon mode can be
large and is highly sensitive to the parameters of the electron spectrum.

\textbf{Note added by S.C. and R.C.} With our deepest regret we must report
that Prof. Maria Marinaro passed away after this work has been completed.

\begin{acknowledgements}
The authors would like to thank Prof. N.L. Saini for the comments.
\end{acknowledgements}


\begin{thebibliography}{99}
\bibitem{graf} J. Graf, M. d'Astuto, C. Jozwiak, D. R. Garcia, N. L. Saini,
M. Krisch, K. Ikeuchi, A. Q. R. Baron, H. Eisaki and A. Lanzara, Phys. Rev.
Lett. \textbf{100}, 227002 (2008).

\bibitem{astuto} M. d'Astuto, G. Dhalenne, J. Graf, M. Hoesch, P. Giura, M.
Krisch, P. Berthet , A. Lanzara and A. Shukla, Phys. Rev. B \textbf{78},
140511(R) (2008)

\bibitem{Reznik} D. Reznik, G. Sangiovanni, O. Gunnarsson and T. P.
Devereaux, Nature \textbf{455}, E6 (2008).

\bibitem{Braden} M. Braden, L. Pintschovius, T. Uefuji, and K. Yamada, Phys.
Rev. B \textbf{72}, 184517 (2005).

\bibitem{pint} L. Pintschovius, Phys. Status Solidi B \textbf{242},30 (2005).

\bibitem{Misochko} O. V. Misochko, E. Ya. Sherman, N. Umesaki, K. Sakai, S.
Nakashima, Phys. Rev. B \textbf{59}, 11495 (2005).

\bibitem{McQ} R.J. McQueeney et al. Phys. Rev. Lett. \textbf{82}, 628 (1999).

\bibitem{PB} L. Pintschovius and M. Braden, Phys. Rev. B \textbf{60,} R15039
(1999).

\bibitem{Reznik1} D. Reznik, L. Pintschovius, M. Ito, S. Iikubo, M. Sato, H.
Goka, M. Fujita, K. Yamada, G. D. Gu, and J. M. Tranquada, Nature (London)
\textbf{440}, 1170 (2006).

\bibitem{dastuto} J. Graf, M. d'Astuto, P. Giura, A. Shukla, N. L. Saini, A.
Bossak, M. Krisch, S.-W. Cheong, T. Sasagawa, and A. Lanzara, Phys. Rev. B
\textbf{76}, 172507 (2007).

\bibitem{ccm} S. Cojocaru, R. Citro, and M. Marinaro, Phys. Rev. B \textbf{75%
}, 014516 (2007).

\bibitem{basov} D.N. Basov and T. Timusk, Rev. Mod. Phys. \textbf{77}, 721
(2005).

\bibitem{Kaneshita} E. Kaneshita, M. Ichioka and K. Machida, Phys. Rev. Lett.%
\textbf{\ } \textbf{88},115501 (2002).

\bibitem{Sherman} E.Ya. Sherman and C. Ambrosch-Draxl, Eur. Phys. J. B
\textbf{26}, 323 (2002).

\bibitem{Khaliullin} G. Khaliullin and P. Horsch, Phys. Rev. B \textbf{54},
R9600 (1996); P. Horsch and G. Khaliullin, Physica B \textbf{359-361}, 620
(2005).

\bibitem{pringle} D. J. Pringle, G. V. M. Williams, and J. L. Tallon, Phys.
Rev. B \textbf{62}, 12527 - 12533 (2000).

\bibitem{iwasawa} H. Iwasawa, J. F. Douglas, K. Sato, T. Masui, Y. Yoshida,
Z. Sun, H. Eisaki, H. Bando, A. Ino, M. Arita, K. Shimada, H. Namatame, M.
Taniguchi, S. Tajima, S. Uchida, T. Saitoh, D.S. Dessau, Y. Aiura, Phys.
Rev. Lett. \textbf{101}, 157005 (2008).

\bibitem{simonelli} L. Simonelli et al. arXiv:0812.0086 [cond-mat.supr-con].

\bibitem{yildirim} T. Yildirim, O. G\"{u}lseren, J. W. Lynn, C. M. Brown, T.
J. Udovic, Q. Huang, N. Rogado, K. A. Regan, M. A. Hayward, J. S. Slusky, T.
He, M. K. Haas, P. Khalifah, K. Inumaru, and R. J. Cava, Phys. Rev. Lett.%
\textbf{\ 87}, 037001 (2001).

\bibitem{crespi} V. H. Crespi and M. L. Cohen, Phys. Rev. B \textbf{48}, 398
(1993).

\bibitem{falter} C. Falter and G. A. Hoffmann, Phys. Rev. B \textbf{61},
14537 (2000), W. Reichardt, J. Low Temp. Phys., \textbf{105}, 807 (1996).

\bibitem{gun} O Gunnarsson and O Rosch, J. Phys.: Condens. Matter \textbf{20}%
, 043201 (2008).

\bibitem{wang} D. T. Wang,A. G\"{o}bel, J. Zegenhagen, and M. Cardona, Phys.
Rev. B \textbf{56}, 13167 (1997).

\bibitem{gobel} A. Gobel, D.T. Wang, M. Cardona, L.Pintschovius, W.
Reichardt, J. Kulda, N.M. Pyka, K. Itoh, E.E. Haller, Phys. Rev. B \textbf{58%
} 10510 (1998).

\bibitem{allen} P.B. Allen, Phys. Rev. B \textbf{6}, 2577 (1972).
\end{thebibliography}
\end{document}